# Molecular Scale Imaging with a Smooth Superlens

*Pratik Chaturvedi[1], Wei Wu[2], VJ Logeeswaran[3], Zhaoning Yu[2], M. Saif Islam[3], S.Y. Wang[2], R. Stanley Williams[2], & Nicholas Fang[1*]*

[1]Department of Mechanical Science & Engineering, University of Illinois at Urbana-Champaign, 1206 W. Green St., Urbana, IL 61801, USA.

[2]Information & Quantum Systems Lab, Hewlett-Packard Laboratories, 1501 Page Mill Rd, MS 1123, Palo Alto, CA 94304, USA.

[3]Department of Electrical & Computer Engineering, Kemper Hall, University of California at Davis, One Shields Ave, Davis, CA 95616, USA.

* Corresponding author Email: nicfang@illinois.edu

**Abstract**

We demonstrate a smooth and low loss silver (Ag) optical superlens capable of resolving features at $1/12^{th}$ of the illumination wavelength with high fidelity. This is made possible by utilizing state-of-the-art nanoimprint technology and intermediate wetting layer of germanium (Ge) for the growth of flat silver films with surface roughness at sub-nanometer scales. Our measurement of the resolved lines of 30nm half-pitch shows a full-width at half-maximum better than 37nm, in excellent agreement with theoretical predictions. The development of this unique optical superlens lead promise to parallel imaging and nanofabrication in a single snapshot, a feat that are not yet available with other nanoscale imaging techniques such as atomic force microscope or scanning electron microscope.

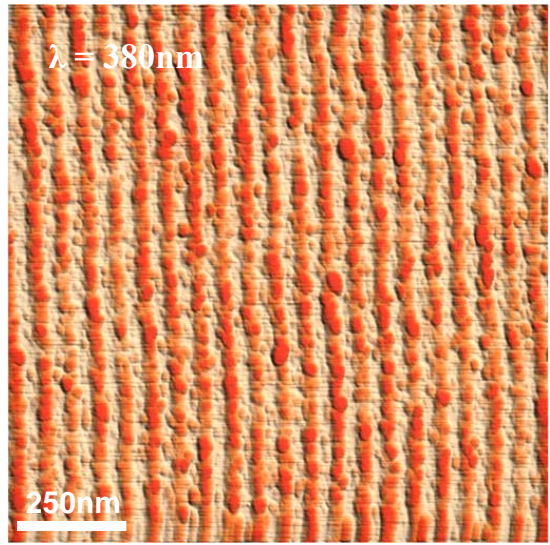

The resolution of optical images has historically been constrained by the wavelength of light, a well known physical law which is termed as the diffraction limit. Conventional optical imaging is only capable of focusing the propagating components from the source. The evanescent components which carry the subwavelength information exponentially decay in a medium with positive permittivity ($\varepsilon$), and positive permeability ($\mu$) and hence, are lost before making it to the image plane. Recent theory suggested a thin negative index film should function as a lens, providing image detail with resolution beyond the diffraction limit.[1] This planar slab of negative index film, termed as "superlens", derives its super-resolution capability from the recovery of evanescent waves via the excitation of surface plasmons. Artificially designed negative index materials, known as metamaterials have demonstrated beam bending[2] and refocusing[3] at microwave frequencies. Achieving negative permeability at optical frequencies is difficult, thus making superlensing extremely challenging. However, in the electrostatic near-field limit, the electric and magnetic responses of materials are decoupled.[1] Thus, for transverse magnetic (TM) polarization, having only negative permittivity suffices for near-field superlensing effect. This makes metals like silver which allow the recovery of evanescent waves,[4] a natural candidate for superlens at optical frequencies. It has been demonstrated experimentally[5] that a silver superlens allows to resolve features well below the working wavelength. Resolution as high as 60nm half-pitch or 1/6$^{th}$ of wavelength has been achieved.

Theoretically, it was predicted that a resolution as high as $\lambda/20$ (where $\lambda$ is the illumination wavelength) is feasible with careful design of silver superlens.[6, 7] However, challenges remain to realize such a high resolution imaging system, such as minimizing the information loss due to evanescent decay, absorption or scattering. Our numerical simulations have indicated that the thickness of spacer layer (separating the object and the lens) and that of silver film are the two major governing factors that determine subwavelength information loss due to evanescent decay and material absorption. Particularly, the surface morphology of silver film plays a significant role in determining the image resolution capability. Below a critical thickness silver is known to form rough islandized films.[8] Rougher films perturb the surface plasmon modes causing loss of subwavelength details and hence diminished resolution.[9] Therefore, producing thin, uniform, and ultra-smooth silver films has been a holy-grail for plasmonics, molecular electronics and nanophotonics. Some recent research efforts directed toward smoothing thin silver films (~100nm thick) have utilized postprocessing steps such as chemical polishing or mechanical pressing.[10] However, the technique suffers from issues common to contact processes such as creation of surface defects, scratches, and delamination of Ag films.

In this work, we utilize a new approach to grow ultra-smooth silver films characterized by much smaller root mean square (RMS) surface roughness. An intermediate ultra-thin Ge layer (~1nm) is introduced before depositing Ag.[11] Utilizing Ag-Ge surface interactions, smooth superlens down to 15nm Ag thickness was fabricated. It is observed that introducing the Ge layer drastically improves Ag surface morphology and helps minimize the island cluster formation. Roughness measurements of thin silver films (15nm) deposited with and without Ge layer (1nm) were performed using atomic force microscopy (AFM) and X-ray reflectivity (XRR) techniques. AFM measurements directly reveal the surface topography and it is observed that the RMS roughness of Ag (over 1x1 μm scan area) deposited on quartz substrate improves from 2.7nm down to 0.8nm by introducing Ge (Figure 1a, b). In XRR measurements, the decay in overall reflected intensity and the oscillation amplitude is strongly affected by the roughness of films. These measurements also suggest drastic improvements in the quality of Ag films incorporating Ge (Figure 1c). Intensity reflected from sample 1 (without Ge - blue dotted curve) drops sharply and does not show oscillating fringes owing to large roughness of the Ag film. In contrast, sample 2 (with Ge – black solid curve) shows large number of fringes and a slow decay in intensity suggesting highly uniform films. Experimental data fit (red dashed curve) reveals that the roughness of 15nm thick Ag film in sample 2 is <0.58nm, more than 4 times smoother compared to sample 1. It is postulated that Ge acts as a wetting layer for Ag and helps a layer by layer growth. A detailed study of the growth of Ag on Ge has been described elsewhere.[11] Some earlier works have indicated that thin and smooth Ag films can also be prepared epitaxially with metal oxides such as magnesium oxide and nickel oxide.[12]

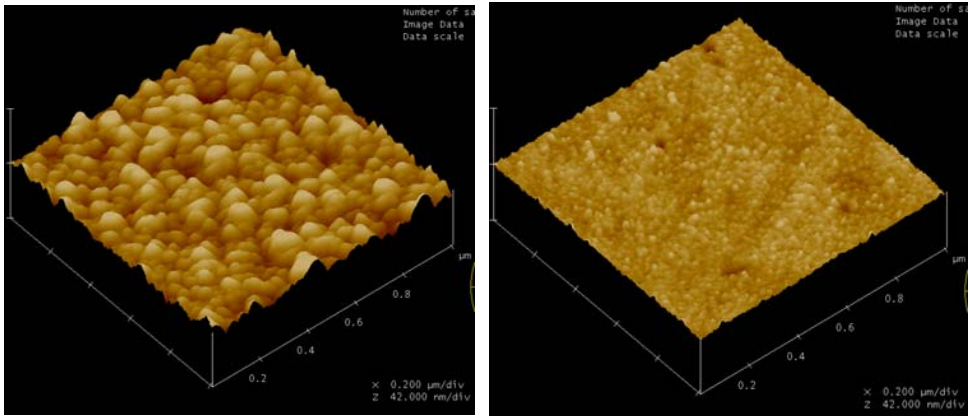

(a)            (b)

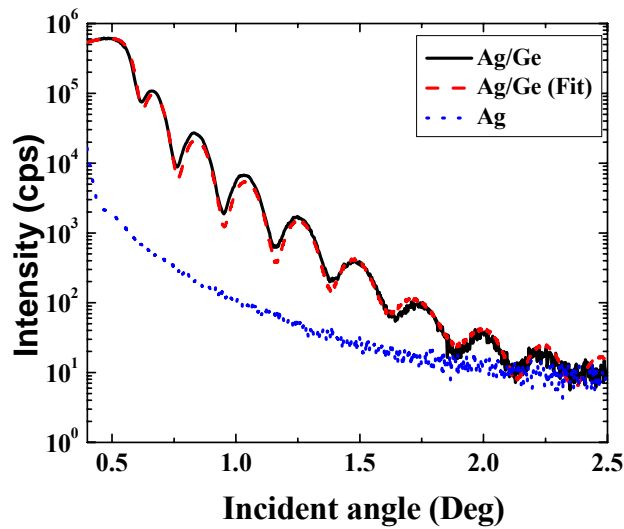

(c)

Figure 1: Smooth Ag growth on Ge (a) Surface topology from AFM micrographs (1x1µm scan) of 15nm Ag deposited on quartz substrate and (b) 15nm Ag deposited

with 1nm Ge on quartz substrate. (c) XRR studies of thin Ag films grown directly on quartz and with Ge intermediate layer.

The configuration of the smooth silver superlens is illustrated in Figure 2. An array of chrome (Cr) gratings 40nm thick with 30nm half-pitch, which serve as the object, was patterned using a nanoimprint process developed by Hewlett-Packard Laboratory. In Figure 3, we present a step-by-step surface characterization of the prepared smooth silver superlens with embedded chrome gratings using AFM. In order to prepare a flat superlens on top of the objects (Figure 3a), it is necessary to deposit a planarization layer to reduce the surface modulations. Surface modulations can alter the dispersion characteristics of the plasmons and it smears out the image details. Also, the planarization layer should be thin to prevent a significant loss of evanescent components from the object. In our process, a planarization procedure using nanoimprint technique is developed to reduce the surface modulations below 1.3 nm (Figure 3b). This is achieved by flood-exposure of 66nm thick UV spacer layer over a flat quartz window under pressure, followed by subsequent reactive ion etching (RIE) to back etch the spacer to 6nm thickness on top of the chrome gratings. A 35nm thick Cr window layer is photolithographically patterned on top of the spacer layer to enhance the contrast with dark-field imaging. Subsequently, 1nm Ge and 15nm Ag layer (superlens) is evaporated over the window layer (Figure 3c), followed by coating with a thick layer of optical adhesive (NOA-73) which serves as the photoresist. The substrate is exposed with a collimated 380nm UV light for 120 seconds (Nichia UV-LED, 80mW). The optical image recorded on the photoresist is developed and imaged with AFM (Figure 3d).

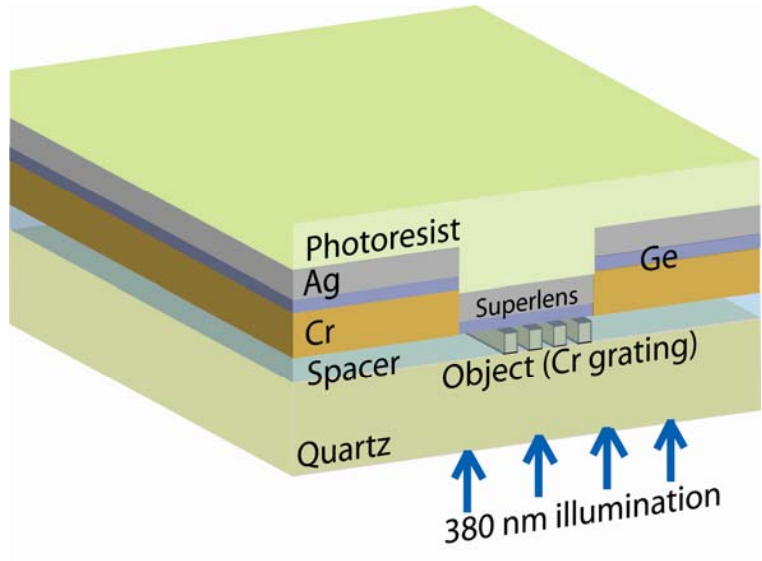

Figure 2: Schematic drawing of a smooth silver superlens with embedded 30nm chrome gratings on a quartz window, operating at 380nm wavelength. To prepare the smooth superlens, a thin germanium layer is seeded.

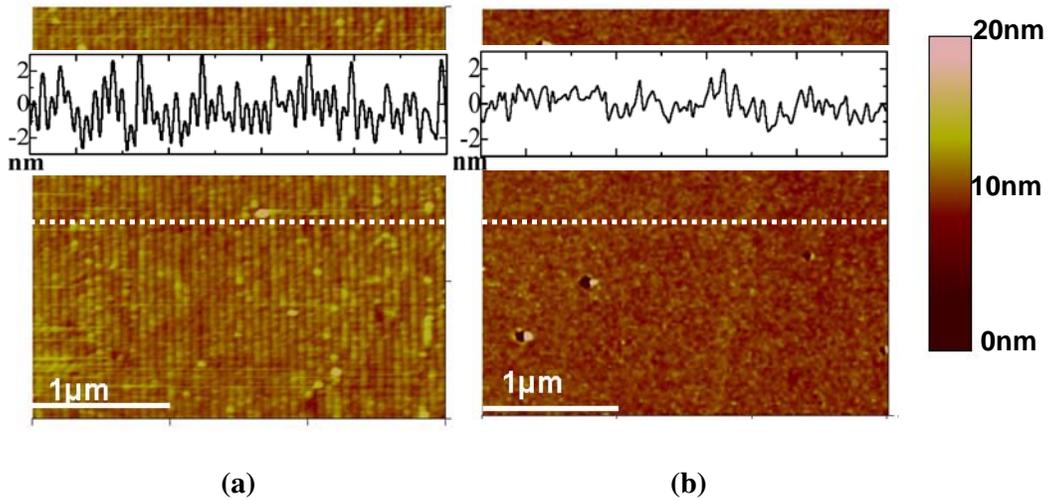

(a)  (b)

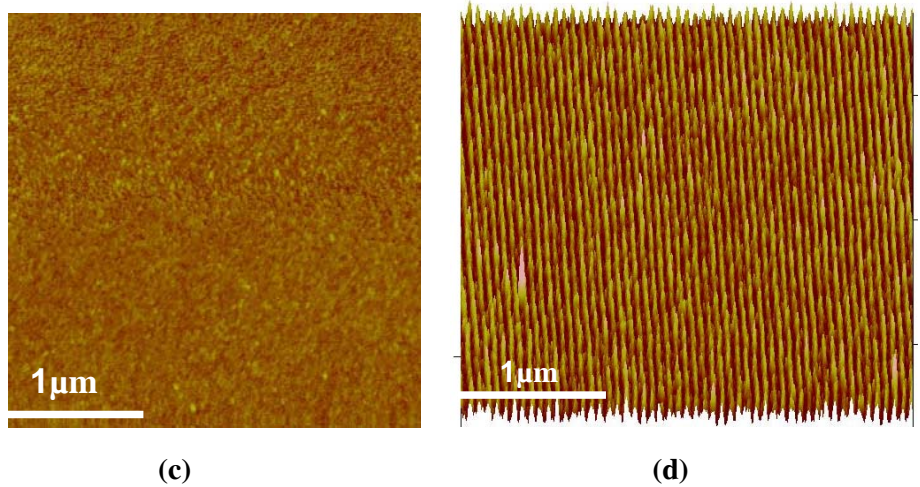

**(c)** **(d)**

Figure 3. Step by step surface characterization of the prepared smooth silver superlens sample with embedded gratings using atomic force microscopy. (a) Close-up image (3X3μm) of the nanoimprinted chrome gratings of 30nm half-pitch prepared on quartz windows. Inset presents the line section plot at the marked dotted line. (b) Surface profile of the sample after planarization with 6nm spacer layer onto chrome gratings, showing an RMS roughness of 1.3nm. (c) Surface profile of the sample after the deposition of Cr window, Ge and Ag layer. (d) The image of the 30nm half-pitch Cr grating area recorded on the photoresist layer after exposure and development**.** (Color scale for all images: 0 to 20nm).

For a qualitative comparison, we theoretically compute the resolving power of a thin ultra-smooth Ag-Ge superlens. Using transfer matrix approach,[13] we compute the optical transfer function and point spread function (PSF) of the multilayer lens system comprising of the spacer(6nm), Ge(1nm) and Ag layers for transverse magnetic polarization at incident wavelength of $\lambda$ = 380nm. We optimize Ag thickness for maximum resolution. It is observed that 20nm thick Ag is capable of transferring a broad range of strongly evanescent modes and can exceed $\lambda$/11 half-pitch resolution.

Adding Ge is generally unfavorable at UV wavelengths, as it is absorptive. However with only 1nm thick Ge in Ag-Ge superlens, the evanescent decay is only significant for feature sizes below λ/12. Computed PSF of such a superlens has full-width at half-maximum (FWHM) of 23nm. An object grating constructed with FWHM of 30nm at 60nm pitch when convoluted with this PSF gives an image grating with FWHM = 37nm (Figure 4a). Moreover, the intensity contrast appearing in the image ($r = \frac{I_{max}}{I_{min}} \sim 3$) is sufficient to resolve this object with most commercial photoresists using superlens photolithography. In contrast, a near-field lens without Ag layer (e.g. spacer 27nm thick) gives a PSF with FWHM of 45nm. Constructed image of the object grating with this lens gives a FWHM of 113nm (~λ /3) (Figure 4b). The resulting image contrast ($r \sim 1.3$) is not sufficient to resolve the grating using photolithography.[14] We experimentally verify our findings by imaging Cr gratings with 30nm width at 60nm pitch using Ag-Ge superlens and near-field control lithography experiments without Ag.

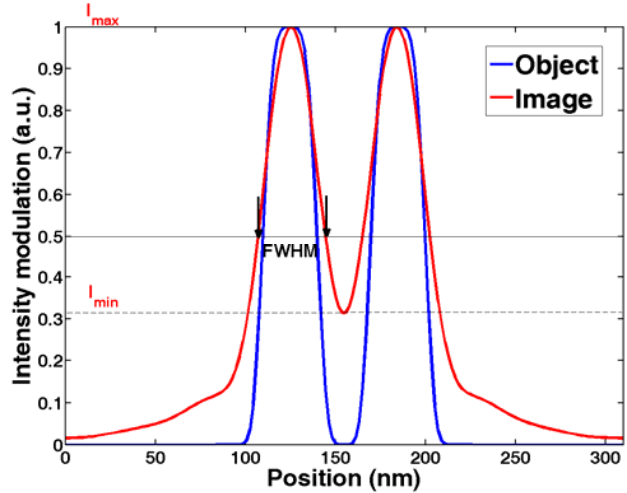

(a)

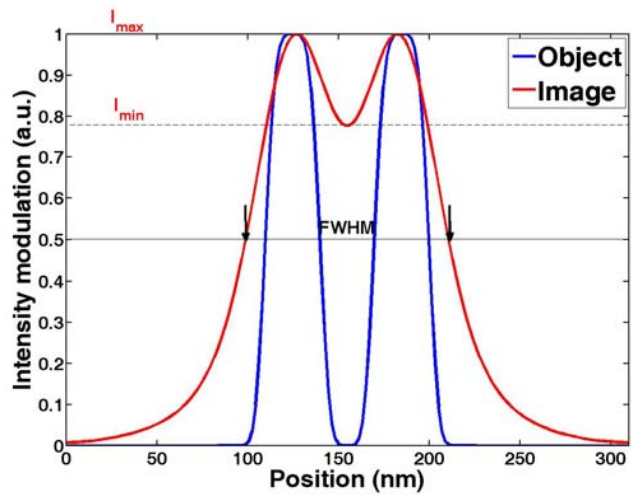

**(b)**

Figure 4: Computed image modulation (a) with superlens FWHM = 37nm (b) without superlens FWHM = 113nm.

Figure 5a (top panel) shows the image of the Cr grating area recorded on the photoresist layer after exposure and development. It is evident from section analysis of the recorded image (middle panel) that with careful control of surface morphology, the recorded image has ~6nm height modulations. The Fourier-transformed spectrum shows clear peaks upto third harmonic of the 60nm pitch Cr gratings successfully recorded on the resist layer (bottom panel). In a control experiment, when the Ag-Ge layers are replaced by equally thick spacer layer, we observe that only a portion of grating area is developed (Figure 5b). Moreover, the developed wires are much thicker (~47nm) and the poor contrast suggests loss of resolution as predicted by the PSF calculation. This confirms that near-field imaging alone without evanescent enhancement is not capable of resolving high-frequency spatial features (~$\lambda/12$) located just 27nm (= $\lambda/14$) away from the surface.

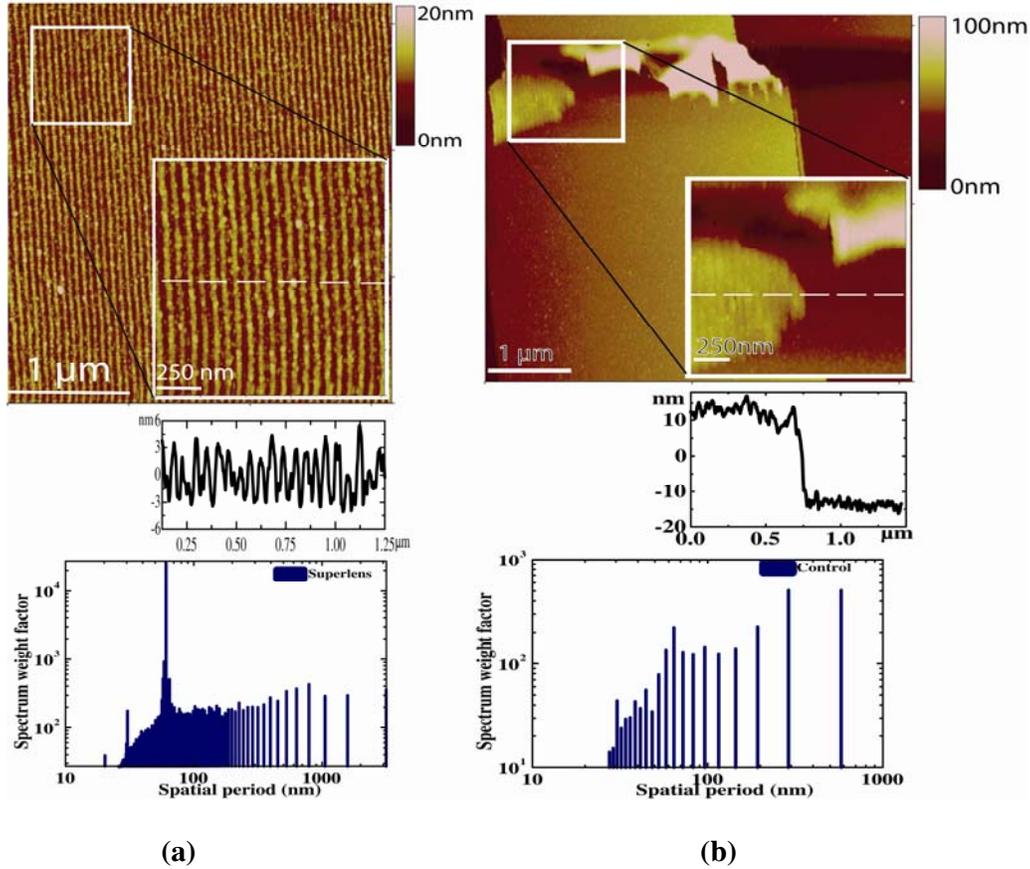

(a) (b)

Figure 5: Subdiffraction optical imaging (a) with superlens (b) without superlens. Top panel: AFM of developed photoresist. Middle panel: Section analysis. Bottom panel: Fourier analysis.

In conclusion, we have demonstrated a new approach to realize ultra-smooth Ag superlenses with an unprecedented $\lambda/12$ resolution capability. Incorporating few monolayers of Ge drastically improves Ag film quality and minimizes the subwavelength information loss due to scattering. Our theoretical and experimental results clearly indicate subdiffraction imaging down to 30nm half-pitch resolution with 380nm illumination. This ultra-high image resolution capability can also be extended to far-field[15, 16] by incorporating a corrugated silver surface on top of Ag-Ge superlens.

Fabrication of sub-20nm thick smooth Ag films will also enable development of novel multilayer (Ag-dielectric-Ag) superlenses operating at visible wavelengths. The development of visible superlenses will allow the use of white light sources instead of specialized UV sources and facilitate possible integration of superlens with optical microscopes. The development of such novel superlenses would enable real-time, dynamic imaging at the molecular level.

**Acknowledgement**

The authors thank Anil Kumar of University of Illinois for help with XRR measurements, Prof. Xiang Zhang of University of California at Berkeley for valuable discussions. XRR experiments were carried out in the Frederick Seitz Materials Research Laboratory Central Facilities, University of Illinois, which are partially supported by the U.S. Department of Energy under grants DE-FG02-07ER46453 and DE-FG02-07ER46471. The authors are grateful for the financial support from the Defense Advanced Research Projects Agency (grant HR0011-05-3-0002), Office of Naval Research (grant N00173-07-G013) and National Science Foundation (grant CMMI-0709023).

**Methods**

*Sample preparation:* Ag and Ag/Ge samples were prepared on quartz using electron-beam evaporation with deposition rate of 0.1 Å/s for Ge and 1 Å/s for Ag, at a pressure of 8E-7 torr. The substrates (~1"x1", 1mm thick) were first cleaned using RCA1 solution.[17] Electronic grade source material was supplied by Kurt J. Lesker with a four-nine purity.

*Nanoimprint technology:* Nanoimprint technology developed at Hewlett-Packard laboratory was utilized to fabricate 30nm half-pitch Cr gratings and 6nm thick spacer films. First a PMMA (950k/15k) layer with thickness 60nm is spin coated on quartz substrate, followed by spin coating of a UV-resist with thickness 66nm. The UV-resist is imprinted and cured using a mold with 30nm half-pitch gratings. UV-resist layer is then etched to 45nm thickness using RIE (with $CF_4$: 60 sccm, 2mtorr), followed by through etching of PMMA layer with $O_2$ RIE (40 sccm, 2mtorr). 40nm thick Cr is deposited using e-beam evaporator at 0.1Å/s, followed by liftoff using acetone and ultrasonic agitation. Planarization of the Cr gratings is performed by spin coating UV-resist with thickness 66nm, followed by imprinting with a flat mold and UV-curing. UV-resist is then etched using $CF_4$ RIE to a total thickness of 46nm, thus resulting in a spacer layer with 6nm thickness.

*XRR measurements:* X-ray reflectivity measurements were carried out on a Philip MRD X'pert system. Measurements were made in a range between 0 and 3 degrees, of which some data points close to zero degree were removed since no useful information is available until the total angle of reflection. Incident angle scan data points were collected with a step width of 0.01 degrees. Theoretical curves were simulated using commercial software Wingixa. Film thickness and density were determined from the period of intensity oscillations and total reflection edge, respectively.